\documentclass[a4paper,11pt]{article}
\pdfoutput=1 % if your are submitting a pdflatex (i.e. if you have
\usepackage{jcappub} % for details on the use of the package, please
\usepackage[T1]{fontenc} % if needed

\usepackage{soul}

\usepackage{aas_macros}

\newcommand{\Mvir}{M_\mathrm{vir}}
\newcommand{\rvir}{r_\mathrm{vir}}
\newcommand{\cvir}{c_\mathrm{vir}}

\title{\boldmath Constraining decaying very heavy dark matter from galaxy clusters with 14 year Fermi-LAT data}

\author[a]{Deheng Song}
\author[b,c,d,e,a]{Kohta Murase}
\author[f,g]{Ali Kheirandish}

\affiliation[a]{Center for Gravitational Physics and Quantum Information,
Yukawa Institute for Theoretical Physics, Kyoto University, Kyoto 606-8502, Japan}
\affiliation[b]{Department of Physics, The Pennsylvania State University, University Park, Pennsylvania 16802, USA}
\affiliation[c]{Department of Astronomy and Astrophysics, The Pennsylvania State University, University Park, Pennsylvania 16802, USA}
\affiliation[d]{Center for Multimessenger Astrophysics, The Pennsylvania State University, University Park, Pennsylvania 16802, USA}
\affiliation[e]{School of Natural Sciences, Institute for Advanced Study, Princeton, New Jersey
08540, USA}
\affiliation[f]{Department of Physics \& Astronomy, University of Nevada, Las Vegas, NV, 89154, USA}
\affiliation[g]{Nevada Center for Astrophysics, University of Nevada, Las Vegas, NV 89154, USA}
\emailAdd{songdeheng@yukawa.kyoto-u.ac.jp}
\emailAdd{murase@psu.edu}
\emailAdd{ali.kheirandish@unlv.edu}

\abstract{
Galaxy clusters are promising targets for indirect detection of dark matter thanks to the large dark matter content. Using 14 years of {\it Fermi}-LAT data from seven nearby galaxy clusters, we obtain constraints on the lifetime of decaying very heavy dark matter particles with masses ranging from $10^3$ GeV to $10^{16}$ GeV. We consider a variety of decaying channels and calculate prompt gamma rays and electrons/positrons from the dark matter. Furthermore, we take into account electromagnetic cascades induced by the primary gamma rays and electrons/positrons, and search for the resulting gamma-ray signals from the directions of the galaxy clusters. We adopt a Navarro-Frenk-White profile of the dark matter halos, and use the profile likelihood method to set lower limits on the dark matter lifetime at a 95\% confidence level. Our results are competitive with those obtained through other gamma-ray observations of galaxy clusters and provide complementary constraints to existing indirect searches for decaying very heavy dark matter.
}

\begin{document}
\maketitle
\flushbottom

\section{Introduction}\label{sec:intro}
Dark matter constitutes the majority of matter in the Universe~\cite{Planck:2018vyg}. Nevertheless, we are still unaware of the particle nature of dark matter. Weakly Interacting Massive Particles (WIMPs) are among the most widely accepted candidates for dark matter~\cite{Bertone:2004pz}. However, 
direct detection experiments have placed significant constraints on WIMPs across a range of masses and they are particularly sensitive to those with masses between approximately 0.01-1 TeV. For WIMPs with masses above 1 TeV, the parameter space remains less constrained, due in part to the lower expected interaction rates.
%direct detection has set very strict constraints for WIMPs with masses below approximately 1 TeV~\cite{Schumann:2019eaa}. Above $\sim1$ TeV, the parameter space of dark matter is relatively unexplored. 
For thermal relics, the unitarity bounds on the WIMP annihilation cross section set an upper limit of dark matter mass at around 100 TeV~\cite{Griest:1989wd}. However, more massive dark matter models are allowed especially if the dark matter is produced non-thermally, for instance, when a period of early matter domination occurs before the Big Bang Nucleosynthesis~\cite{Bramante:2017obj, Cirelli:2018iax, Bhatia:2020itt}, or when dark matter is produced in a hidden sector~\cite{Dubrovich:2003jg, Hisano:2006nn, vonHarling:2014kha, Baldes:2017gzw, Smirnov:2019ngs, Baldes:2021aph}.

Indirect detection, which looks for standard model particles originating from dark matter in the cosmos, is a viable approach for searching dark matter. Thermally produced dark matter particles can still self-annihilate to Standard Model particles in the Universe today. They can also have finite lifetimes, and decay into Standard Model products. High-energy gamma-ray and neutrino telescopes can detect gamma rays and neutrinos produced through dark matter annihilation or decay, which can travel long distances in the Universe. The {\it Fermi} Large Area Telescope ({\it Fermi}-LAT) has been extensively used to search for indirect signals from dark matter. Towards the Galactic center, {\it Fermi}-LAT has identified an excess in the GeV energies, which may be caused by WIMP dark matter annihilation or a population of unresolved millisecond pulsars~\cite{Goodenough:2009gk, Hooper:2011ti, Hooper:2010mq, Abazajian:2012pn, Gordon:2013vta, Abazajian:2014fta, Calore:2014xka, Daylan:2014rsa, Fermi-LAT:2015sau}. Although the origin of the gamma-ray excess is still debated, the Galactic center region still places {meaningful} constraints on the dark matter annihilation cross section~\cite{Hooper:2012sr}. {\it Fermi}-LAT is also used to search for dark matter in the range of $\sim$ GeV to TeV from dwarf galaxies~\cite{Cholis:2012am, Carlson:2014nra, MAGIC:2016xys, Lopez:2015uma, Fermi-LAT:2015att, Baring:2015sza, Calore:2018sdx, Hoof:2018hyn, Alvarez:2020cmw, Gammaldi:2021zdm}, galaxy clusters~\cite{2010JCAP...05..025A, Huang:2011xr, 2012JCAP...07..017A, Murase:2012rd, Fermi-LAT:2015xij, Tan:2019gmb, Thorpe-Morgan:2020czg, DiMauro:2023qat}, and the diffuse isotropic gamma-ray background~\cite{Cirelli:2009dv, Abazajian:2010zb, Murase:2012xs, Bringmann:2013ruh, Ajello:2015mfa, DiMauro:2015tfa, Fermi-LAT:2015qzw}. Galaxy clusters are the largest gravitationally-bound systems in the Universe and host enormous dark matter halos. TeV gamma-ray telescopes, such as the Very Energetic Radiation Imaging Telescope Array System (VERITAS), the Major Atmospheric Gamma Imaging Cherenkov (MAGIC) and the High Altitude Water Cherenkov (HAWC) observatory, have searched for dark matter signals from different galaxy clusters and have placed stringent limits on dark matter annihilation and decay~\cite{2010ApJ...710..634A, 2012ApJ...757..123A, Palacio:2015nza, Harding:2015bua, MAGIC:2018tuz, HAWC:2021udn}. Future TeV gamma-ray telescopes are expected to extend the search for heavy dark matter.~\cite{Maity:2021umk, Tak:2022vkb}.

Dedicated analyses of multi-messenger astrophysical data have set strongest constraints on the lifetime of very heavy dark matter (VHDM) with masses up to EeV or even the GUT (Grand Unified Theory) scale ($\sim$10$^{16}$ GeV)~\cite{Murase:2012xs, Ishiwata:2019aet, Chianese:2021jke, Chianese:2021htv,Das:2023wtk,Fiorillo:2023clw}. 
The IceCube Neutrino Observatory~\cite{IceCube:2018tkk, Arguelles:2019boy, IceCube:2021sog} and the large high altitude air shower observatory (LHAASO)~\cite{LHAASO:2022yxw} also searched for decaying dark matter with masses above PeV and set competitive limits. Future neutrino experiments will extend the search for VHDM to $\gtrsim 10^{12}$ GeV \cite{Arguelles:2022nbl}.

In this work, we search for signals from decaying VHDM in galaxy clusters using 14 year gamma-ray data from {\it Fermi}-LAT. We select seven nearby galaxy clusters based on their dark matter content~\cite{2018PhRvL.120j1101L}. We examine a wide range of dark matter masses, from $\sim 10^3$ GeV to the GUT scale ($\sim 10^{16}$ GeV). As {\it Fermi}-LAT is only sensitive to gamma rays with energies up to $\sim$ TeV, it cannot detect prompt gamma rays from most of the mass range we consider. However, the high-energy gamma rays and electrons/positrons ($e^\pm$) produced by dark matter decay interact with the radiation and magnetic fields inside galaxy clusters, causing electromagnetic cascades. The secondary gamma rays generated through inverse-Compton and synchrotron processes are cascaded down to the energy range that {\it Fermi}-LAT can detect~\cite{Murase:2012xs, Murase:2012rd, Murase:2015gea, Cohen:2016uyg, Blanco:2018esa}. We search for gamma-ray signals from dark matter decay in galaxy clusters taking into account the electromagnetic cascades and place limits on the lifetime of dark matter.

The rest of this paper is organized as follows. In section~\ref{sec:signal}, we describe the expected signals from VHDM decay, including the electromagnetic cascades. In section~\ref{sec:clusters}, we summarize the selection of galaxy clusters. We provide the details about our data analysis procedures in section~\ref{sec:data}. Subsequently, we present and discuss our results in section~\ref{sec:results} and summarize the paper in section~\ref{sec:summary}.

\section{Signals of VHDM decay}\label{sec:signal}
The expected intensity of generated gamma rays from decaying dark matter for given gamma-ray energy $E_\gamma=E'_\gamma/(1+z)$ and direction $\Omega$ on the sky is
\begin{equation}
    I_{\gamma}^{(\rm gen)}(E_\gamma,\Omega) = \frac{1}{4\pi}\frac{1}{\tau_\chi m_\chi}\frac{dS_\gamma}{dE'_\gamma} 
    \int dl \rho_\chi(r),
\end{equation}
where $\tau_\chi$ and $m_\chi$ are the lifetime and mass of dark matter particle $\chi$, and ${dS_\gamma}/{dE'_\gamma}$ is the gamma-ray spectrum from dark matter decay to Standard Model particles, which is normalized via
\begin{equation}
\Sigma_i \int d E'_i E'_i \frac{dS_i}{dE'_i}=m_\chi,
\end{equation}
where the summation is for all particle species. Here, $\rho_\chi$ is the smooth density distribution of dark matter, which is integrated over the solid angle $\Omega$ and the comoving line-of-sight distance $l$. For dark matter halos of galaxy clusters, we use a Navarro–Frenk–White (NFW) profile~\cite{1997ApJ...490..493N} in comoving coordinates as the following
\begin{equation}
    \rho_\chi(r) = \frac{\rho_s}{(r/r_s)(1+r/r_s)^2},
\end{equation}
where the scale density $\rho_s$ and scale radius $r_s$ vary from cluster to cluster. 

The generated gamma-ray energy flux within a cone with a sold angle of $\Omega$ is given by
\begin{equation}
    E_\gamma F_{E_\gamma}^{(\rm gen)}(\Omega) = \int d\Omega E_\gamma^2 I_{\gamma}^{(\rm gen)}(\Omega) = \frac{1}{4\pi d_L^2}\frac{1}{\tau_\chi m_\chi}{E'}_\gamma^2\frac{dS_\gamma}{dE'_\gamma}\times \frac{4\pi d_c^2}{4\pi} J^{\mathrm{dec}},
\label{eq:DMflux}
\end{equation}
where $d_L$ and $d_c$ are luminosity distance and comoving distance, respectively, and the angle-integrated J-factor $J^{\mathrm{dec}}(\Omega)$ for decaying dark matter is given by~\cite{2018PhRvL.120j1101L}
\begin{equation}
    J^{\mathrm{dec}}(\Omega)= \int d\Omega \int dl \rho_\chi(r)=\int d\Omega {\mathcal J}^{\rm dec}(\Omega),
\end{equation}
where ${\mathcal J}^{\rm dec}$ is the angle-dependent J-factor of the cluster. In this work, we use the public software \texttt{CLUMPY}~\cite{2012CoPhC.183..656C, 2016CoPhC.200..336B, 2019CoPhC.235..336H} to calculate ${\mathcal J}^{\rm dec}$ following the NFW profile. Note that the integration over the virial radius gives $J^{\rm dec}\approx M_{\rm vir}/d_c^2$, where $M_{\rm vir}$ is the virial radius, and the normalized J-factor becomes~\cite{Murase:2012rd}
\begin{equation}
    {\hat{J}}^{\mathrm{dec}}(\theta)=\frac{2\pi \int d\theta \theta E_\gamma^2 I_\gamma}{E_\gamma F_{E_\gamma}}\approx \frac{d_c^2}{M_{\rm vir}}J^{\mathrm{dec}}(\theta),
\end{equation}
where $\theta=\tan^{-1}(r/d_c)$. 

In this study, we consider VHDM in the mass range between $10^3$ GeV to $10^{16}$ GeV assuming that each dark matter particle decays into a pair of Standard Model particles. We consider seven decaying channels: $b\Bar{b}$, $t\Bar{t}$, $e^+e^-$, $\mu^+\mu^-$, $\tau^+\tau^-$, $Z^0Z^0$, and $W^+W^-$. To calculate the prompt spectra, we use \texttt{HDMSpectra}~\cite{2021JHEP...06..121B}, which incorporates all relevant electroweak interactions. This contribution is crucial for VHDM. 

Equation~\ref{eq:DMflux} is for the flux of generated gamma rays, but for gamma rays from VHDM we must further take into account contributions from electromagnetic cascades that are developed inside galaxy clusters following the treatments described in Refs.~\cite{Murase:2012xs,Murase:2012rd}. We solve the Boltzmann equations of photons and $e^\pm$ considering $e^\pm$ pair creation, inverse-Compton scattering and synchrotron radiation. Prompt $e^\pm$ from VHDM decay are trapped in the galaxy clusters and lose energies by interacting with the cosmic microwave background (CMB) and extragalactic background light (EBL) as well as intra-cluster magnetic fields, leading to inverse-Compton and synchrotron emission. When dark matter is sufficiently heavy ($m_\chi \gtrsim$ 10 TeV), prompt gamma rays are attenuated by radiation fields in the galaxy clusters and secondary $e^\pm$ are generated. These secondary $e^\pm$ also produce inverse-Compton and synchrotron emissions. 
The final gamma-ray spectra from dark matter decay in galaxy clusters combine the remaining prompt and the total secondary gamma rays. {We include the details about intra-cluster electromagnetic cascades in Appendix~\ref{sec:cascades}.}
After the gamma rays leave the galaxy clusters and enter the intergalactic space, they are further attenuated by the EBL. We calculate such attenuation, but {ignore intergalactic cascade emission as in previous works~\cite{Murase:2012rd}. This is because they are expected to form a giant pair halo with a size of $\lambda_{\gamma\gamma}/d$~\cite{Aharonian:1993vz,Murase:2008pe,Murase:2012df}, where $\lambda_{\gamma\gamma}$ is the mean free path to the two-photon annihilation process and $d$ is the source distance. The mean free path of 100~TeV gamma rays is around $\lambda_{\gamma\gamma}\sim3$~Mpc~\cite{Murase:2011yw}, and higher-energy gamma rays are cascaded inside clusters. The mean free path of 10~TeV gamma rays is $\lambda_{\gamma\gamma}\sim100-200$~Mpc, which is larger than the distance to nearby clusters. 
Therefore most of the intergalactic cascade emission contributes to diffuse or even quasi-isotropic emission, and it is reasonable and more conservative to ignore the contribution for the present analyses. See Ref.~\cite{Murase:2009zz} for the total contribution from a source and extended/diffuse intergalactic cascade emission.}

\begin{figure}[tbp]
    \centering
    \begin{minipage}[t]{0.49\textwidth}
        \centering
        \includegraphics[width=\textwidth]{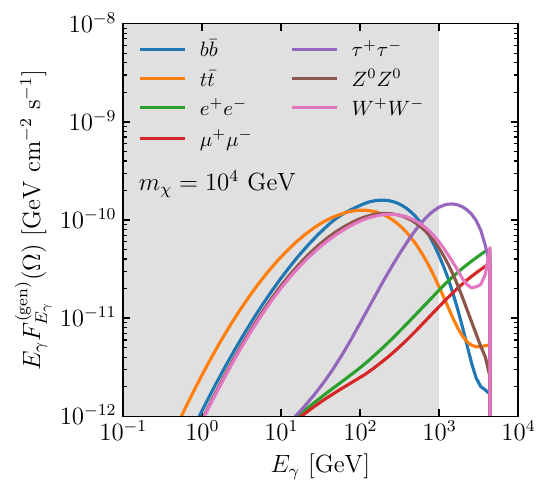}
        \caption{Prompt spectra of gamma rays from dark matter decay. Here, we show the case of $m_\chi = 10^4$ GeV and show seven decaying channels: $b\Bar{b}$, $t\Bar{t}$, $e^+e^-$, $\mu^+\mu^-$, $\tau^+\tau^-$, $Z^0Z^0$, and $W^+W^-$. The grey area shows the {\it Fermi}-LAT energy range. The fluxes are normalized by $J^{\rm dec}(\Omega)$ of Virgo.
        % \km{Please use $E_\gamma F_{E_\gamma}^{(\rm gen)}$ in the caption.}
        }
        \label{fig:spectra_gamma}
    \end{minipage}
    \vspace{1.00mm}
    \begin{minipage}[t]{0.49\textwidth}
        \centering
        \includegraphics[width=\textwidth]{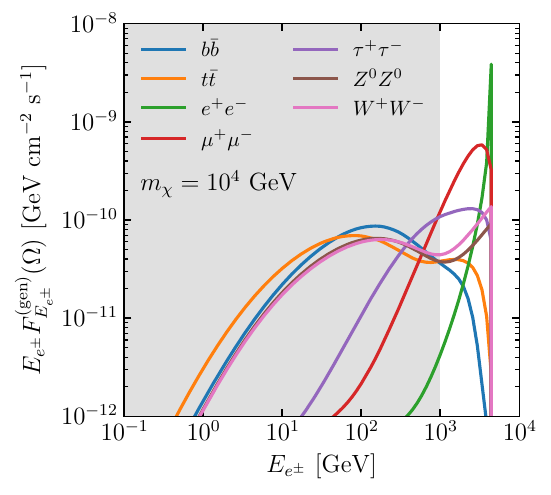}
        \caption{Same as figure~\ref{fig:spectra_gamma}, but of $e^\pm$.
        % \km{Please use $E_\gamma F_{E_\gamma}^{(\rm gen)}$ in the caption.}
        }
        \label{fig:spectra_epm}
    \end{minipage}
\end{figure}

\begin{figure}[tbp]
    \centering
    \begin{minipage}[t]{0.49\textwidth}
        \centering
        \includegraphics[width=\textwidth]{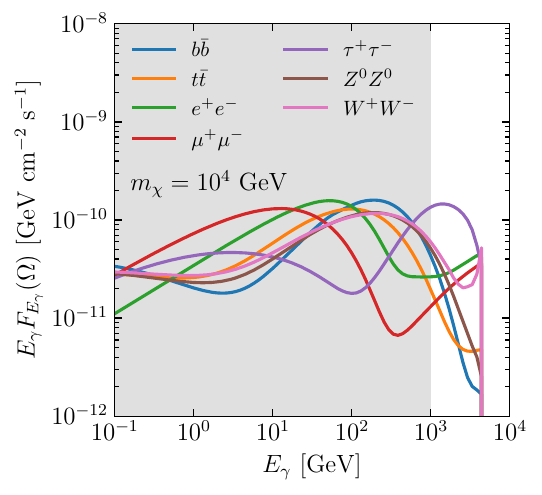}
        \caption{Gamma-ray spectra from dark matter decay when electromagnetic cascades are considered.}
        \label{fig:spectra_cascaded_1e4}
    \end{minipage}
    \vspace{1.00mm}
    \begin{minipage}[t]{0.49\textwidth}
        \centering
        \includegraphics[width=\textwidth]{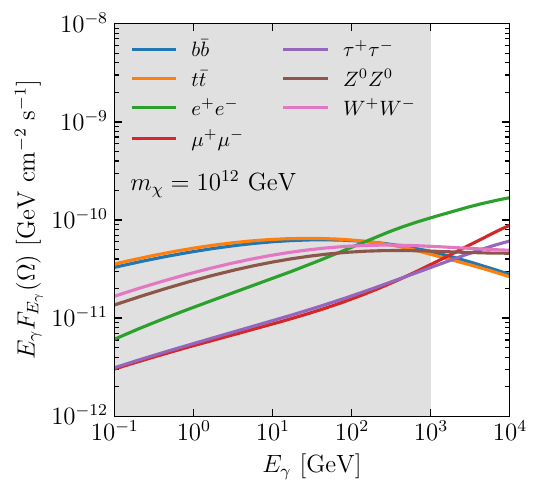}
        \caption{Same as figure~\ref{fig:spectra_cascaded_1e4}, but for $m_\chi = 10^{12}$ GeV.}
        \label{fig:spectra_cascaded_1e12}
    \end{minipage}
\end{figure}

Figure~\ref{fig:spectra_gamma} and figure~\ref{fig:spectra_epm} show the prompt gamma-ray and $e^\pm$ spectra resulting from dark matter decay, respectively. The gamma-ray and $e^\pm$ fluxes are normalized by assuming $\tau_\chi = 10^{28}$~s and $J^{\mathrm{dec}}(\Omega)$ of the Virgo cluster. The grey area denotes the energy range to which {\it Fermi}-LAT is sensitive. These figures assume $m_\chi = 10^4$ GeV and include seven channels. For the $b\Bar{b}$, $t\Bar{t}$, $Z^0Z^0$, and $W^+W^-$ channels, the decaying final states are largely hadronic, resulting in similar gamma-ray spectra at low energies. The gamma-ray spectra of the $e^+e^-$ and $\mu^+\mu^-$ channel is very hard and peaks at $\sim m_\chi/2$ as they are dominated by the final-state radiation. Therefore, including electromagnetic cascades is more critical for the $e^+e^-$ and $\mu^+\mu^-$ channel. The $\tau^+\tau^-$ channel is intermediate since the leptonic and hadronic decays are comparable~\cite{2018PhRvL.120j1101L}. Figure~\ref{fig:spectra_cascaded_1e4} shows the expected gamma-ray spectra from dark matter decay when electromagnetic cascades are considered, for $m_\chi = 10^4$ GeV. We assume that the magnetic field in the cluster is $B_\mathrm{cl} = 0.3\ \mu$G~\cite{2004IJMPD..13.1549G} and the size of the emission region is set to $R_{\rm cl} = 3$ Mpc. We discuss the impact of these parameters later in section~\ref{sec:results}. Comparing to figure~\ref{fig:spectra_gamma}, the spectra including electromagnetic cascades extend to lower energies due to contributions from inverse-Compton and synchrotron radiation processes. The spectra are also undulating when transitioning from prompt to secondary gamma rays. When dark matter is sufficiently heavy, {\it Fermi}-LAT will only be able to detect secondary gamma rays, as demonstrated in figure~\ref{fig:spectra_cascaded_1e12} for $m_\chi = 10^{12}$ GeV. In this case, gamma-ray spectra in the {\it Fermi}-LAT energy range are dominated by synchrotron radiation and are broad and smooth.


\section{Samples of galaxy clusters}\label{sec:clusters}
\begin{table}[tbp]
\centering
\begin{tabular}{|ccccccccc|}
\hline
Cluster & $l$ & $b$ & $z \times 10^3$ & $\cvir$ & $\rvir$ &  $\theta_\mathrm{vir}$ & $\log_{10}\Mvir$& $\log_{10}J^\mathrm{dec.}$ \\
& [deg] & [deg] & &  & [kpc] & [deg] & [$M_\odot$] & [GeV cm$^{-2}$ sr] \\
\hline
Virgo & 283.94 & 74.52 & 3.58 & 6.36 & 2023.32 & 7.28 & 14.66 & 20.41\\
Centaurus & 302.22 & 21.65 & 8.44 & 6.40 & 1971.53 & 3.02 & 14.62 & 19.63\\
Norma & 325.29 & $-$7.21 & 17.07 & 5.60 & 2842.43 & 2.16 & 15.10 & 19.51\\
Perseus & 150.58 & $-$13.26 & 17.62 & 5.63 & 2796.52 & 2.06 & 15.08 & 19.46\\
Coma & 57.20 & 87.89 & 24.45 & 5.40 & 3132.32 & 1.67 & 15.23 & 19.33\\
Hydra & 269.55 & 26.41 & 10.87 & 6.74 & 1687.23 & 2.01 & 14.42 & 19.21\\
Fornax & 239.98 & $-$56.68 & 4.17 & 8.47 & 873.63 & 2.71 & 13.56 & 19.17\\
\hline
\end{tabular}
\caption{\label{tab:clusters} Sample of galaxy clusters. The parameters are taken from Ref.~\cite{2018PhRvL.120j1101L}.}
\end{table}

We choose the galaxy clusters from the catalog presented in Ref.~\cite{2018PhRvL.120j1101L}. This catalog was initially developed based on \cite{2015AJ....149..171T, 2017ApJ...843...16K}. Here, we select seven clusters with the largest $J^\mathrm{dec.}$. Table~\ref{tab:clusters} summarizes the properties of the clusters, including locations, redshifts ($z$), and dark matter halo parameters. The selected clusters have virial masses of $M_\mathrm{vir}\sim 10^{14} - 10^{15} M_\odot$. The virial radius $r_\mathrm{vir}$ and the concentration parameter $\cvir$ are related to the scale radius of the dark matter halo, $r_s = \rvir/\cvir$. We also report the angular size of the dark matter halo at the virial radius, which is given by
\begin{equation}
    \theta_\mathrm{vir} = \tan^{-1}(\rvir/d_c).
\end{equation}
The Virgo cluster has the largest $J^\mathrm{dec.}$, which is $\sim 10^{20}$ GeV cm$^{-2}$ sr, and is the closest cluster with a redshift $z=3.58\times 10^{-3}$. Therefore, the dark matter halo of Virgo extends to $\theta_\mathrm{vir} = 7.28^\circ$. The remaining clusters are: Centaurus, Norma, Perseus, Coma, Hydra, and Fornax. They typically have $J^\mathrm{dec.} \sim 10^{19}$ GeV cm$^{-2}$ sr, and $\theta_\mathrm{vir} \sim 2^\circ-3^\circ$.

\section{Data analysis}\label{sec:data}
We analyze 14 year {\it Fermi}-LAT Pass8 data collected between Aug 4 2008 and Aug 4 2022. We select photon events from the \texttt{ULTRACLEANVETO} class and include both \texttt{FRONT} and \texttt{BACK} types. We apply a quality filter ``\texttt{DATA\_QUAL>0 \&\& LAT\_CONFIG==1}'' and limit the maximum zenith angle to 90$^\circ$. For each galaxy cluster, the region of interest (ROI) is a 20$^\circ$ by 20$^\circ$ square centered at the cluster's location with an angular resolution of 0.1$^\circ$. The energy range of the gamma rays is from 100~MeV to 1~TeV, with 5 logarithmic energy bins per decade.

We fit the {\it Fermi} data using the open-source Python package \texttt{fermipy}~\cite{2017ICRC...35..824W}. For each cluster, we include the Galactic interstellar emission model \texttt{gll\_iem\_v07.fits} and isotropic diffuse template \texttt{iso\_P8R3\_ULTRACLEANVETO\_V3\_v1.txt} as the background models. Also included are the point sources resolved in the LAT 12-year source catalog (4FGL-DR3)~\cite{2020ApJS..247...33A,2022ApJS..260...53A} that are up to 5$^\circ$ away from the boundary of the ROI. Since the 4FGL-DR3 only contains point sources resolved with 12 years of {\it Fermi} data, new point sources may emerge in the 14 year data we use. These new point sources, if they exist, must be included in the background model. Otherwise, their photons may be attributed to dark matter models. We use the \texttt{find\_sources} algorithm provided by \texttt{fermipy} to search for new point sources in each ROI and include them in the background model. We find seven new point sources in the ROIs for four of the galaxy clusters in our list. Table~\ref{tab:add_ps} summarizes these new point source locations, offsets from the galaxy clusters, and test statistic (TS)\footnote{{We note that the \texttt{find\_sources} algorithm provided by \texttt{fermipy} only offers a straightforward routine to search for potential new sources after fitting the known background sources. We include this procedure to optimize our background models for the search of VHDM signals. The true nature of these sources requires further investigation.}}. There sources are included in the background models and their normalizations are allowed to vary.

We consider the gamma-ray signal from dark matter as an extended emission in the ROI up to the virial radius of the dark matter halo. We use the public software \texttt{CLUMPY}~\cite{2012CoPhC.183..656C, 2016CoPhC.200..336B, 2019CoPhC.235..336H} to calculate the angular-dependent J-factor $dJ^\mathrm{dec.}/d\Omega(\phi)$ of the cluster following an NFW profile and use it as a diffuse template in the analysis. {\texttt{CLUMPY} has been widely used to generate templates of dark matter halos that can be directly used in the analysis of \textit{Fermi}-LAT data.} In each fit, we fix $m_\chi$ and the decaying channel of dark matter and the only free parameter of the dark matter model is the normalization of the gamma-ray flux, which is inversely proportional to $\tau_\chi$. The best-fitting model is found by maximizing the Poissonian likelihood function summed over each pixel $i$ and energy bin $j$ for the $k$-th cluster, which is given by
\begin{equation}
    \mathcal{L}^k(\theta) = \prod_{ij}\dfrac{\mu^k_{ij}(\theta)^{n^k_{ij}}e^{-\mu^k_{ij}(\theta)}}{n^k_{ij}!}.
\end{equation}
Here, $n^k_{ij}$ represents the observed photon counts, and $\mu^k_{ij}(\theta)$ represents the predicted photon counts at the pixel $i$ and energy bin $j$ of the $k$-th cluster. The parameter $\theta = \{\tau_\chi, \sum_l\lambda_l\}$ is a set of parameters includes the dark matter lifetime $\tau_\chi$ and the parameters of the astrophysical backgrounds $\lambda_l$. For the backgrounds, we vary the normalizations and spectral shapes of the diffuse emissions and the point sources within 10$^\circ$ from the centers of the ROIs. 

We obtain the limit on $\tau_\chi$ at the 95\% confidence interval (CL) using the profile likelihood method~\cite{Rolke:2004mj, 2012ApJ...761...91A}. For each individual cluster, we scan $\tau_\chi$ and refit the background parameters $\lambda_l$ to find the change in the likelihood function from the best-fitting model. The 95\% CL limit on the dark matter lifetime is set by finding the set of nuisance parameters for which 
\begin{equation}
    -2\Delta\log \mathcal{L}^k= -2\left(\log\mathcal{L}^k(\hat{\theta}) - \log\mathcal{L}^k(\theta) \right) = {-}2.71,
\end{equation}
where $\hat{\theta}$ is the best fit parameters we found for cluster.

\begin{table}[tbp]
\centering
\begin{tabular}{|llcccc|}
\hline
Cluster & Name & $l$ & $b$ & offset & TS \\
 &  & [deg] & [deg] & [deg] &  \\
\hline
Centaurus & PS J1320.6$-$4524 & 308.29 & 17.15 & 7.28 & 68.12\\
Centaurus & PS J1158.2$-$4419 & 293.00 & 17.51 & 9.62 & 30.01\\
Norma & PS J1524.6$-$6320 & 319.16 & $-$5.40 & 6.36 & 33.34\\
Perseus & PS J0254.8+3934 & 147.40 & $-$17.36 & 5.12 & 38.28\\
Perseus & PS J0400.1+3710 & 159.61 & $-$11.91 & 8.92 & 35.55\\
Coma & PS J1323.3+3212 & 71.07 & 81.42 & 6.56 & 45.37\\
Coma & PS J1316.6+2056 & 347.17 & 81.55 & 7.98 & 32.06\\
\hline
\end{tabular}
\caption{\label{tab:add_ps} Additional point sources included in the ROIs.}
\end{table}

\section{Results and discussion}\label{sec:results}
\begin{figure}[tbp]
    \centering
    \begin{minipage}{0.49\textwidth}
        \centering
        \includegraphics[width=\textwidth]{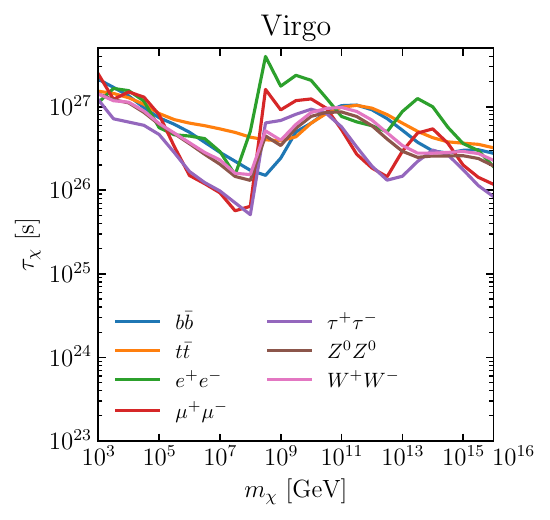}
        \caption{Limits on $\tau_\chi$ for five decaying channels from the Virgo cluster.}
        \label{fig:limits_virgo}
    \end{minipage}
    \vspace{1.00mm}
    \begin{minipage}{0.49\textwidth}
        \centering
        \includegraphics[width=\textwidth]{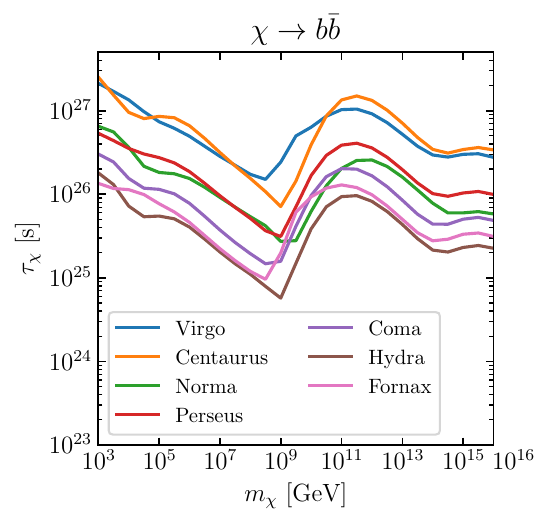}
        \caption{Limits on $\tau_\chi$ for the $b\bar{b}$ channel from seven galaxy clusters.}
        \label{fig:limits_bbar}
    \end{minipage}
\end{figure}

In figure~\ref{fig:limits_virgo}, we present the 95\% CL lower limits on $\tau_\chi$ for five decaying channels from the Virgo cluster for $10^3$ GeV $< m_\chi < 10^{16}$ GeV. Overall, the most stringent limits have $\tau_\chi \gtrsim 10^{27}$s at $m_\chi \sim$ TeV. The limits become less stringent when $m_\chi$ increases since the expected signals change from prompt dominated region to secondary (inverse-Compton) dominated region, which gradually becomes less intense. The limits weaken to $\tau_\chi \gtrsim 10^{26}$ s when $m_\chi$ is around EeV. For $m_\chi \gtrsim$ EeV, synchrotron emission from prompt and secondary $e^\pm$ becomes important in the {\it Fermi} energy range, and the limits tend to recover from $m_\chi\sim$ EeV. For $b\bar{b}$ and $t\bar{t}$ and $Z^0Z^0$ and $W^+W^-$ channels, the limits gradually increase and eventually reach to $\tau_\chi \gtrsim 10^{27}$ at $m_\chi\sim$ 100 EeV. Above $m_\chi\sim$ 100 EeV, the limits gradually become weaker as the peak of the synchrotron emission pass the {\it Fermi} energy range. For $e^+e^-$, $\mu^+\mu^-$ and $\tau^+\tau^-$ channels, the behavior of the limits is different since these channels are largely leptonic. The limits recover early and quickly to $\tau_\chi \gtrsim 10^{27}$ s at $m_\chi\sim$ EeV. The limits also show a second peak at $10^{14}$ GeV since the inverse-Compton photons are attenuated and pair-create $e^\pm$, generating a secondary synchrotron component.

\begin{figure}[tbp]
    \centering
    \includegraphics[width=0.8\columnwidth]{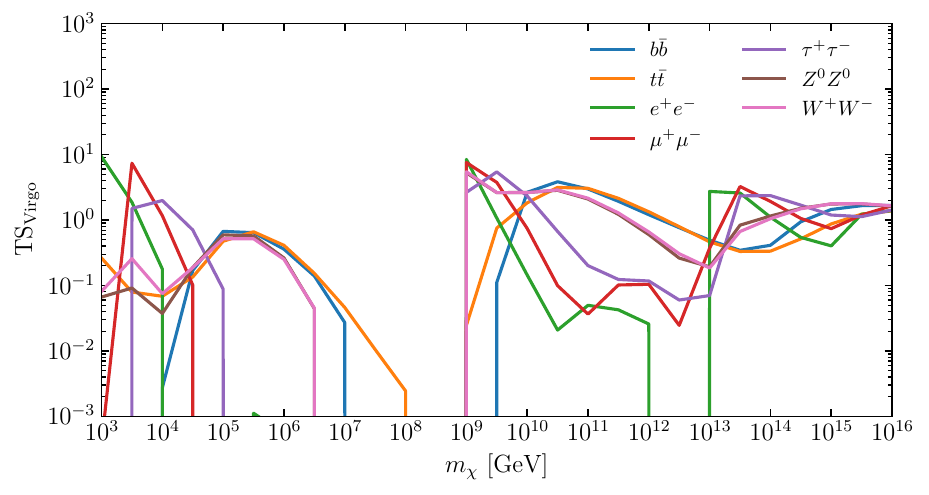}
    \includegraphics[width=0.8\columnwidth]{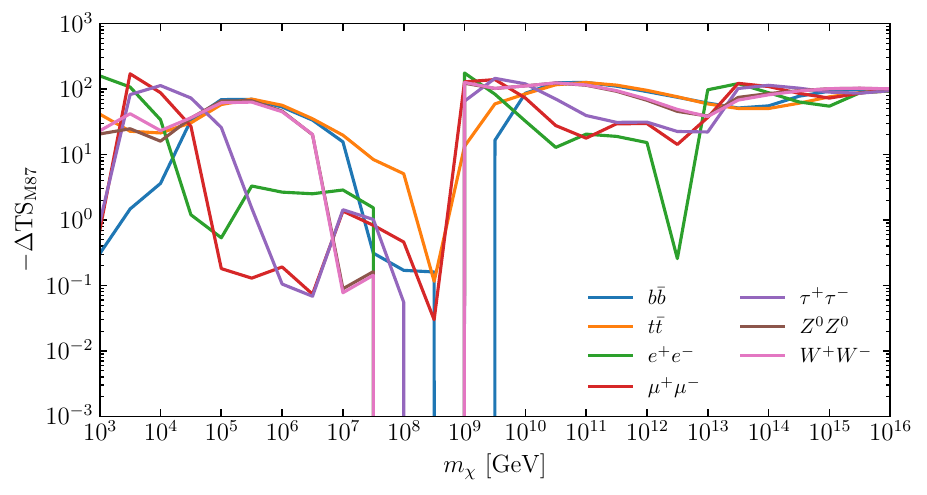}
    \caption{\emph{\textbf{Top.}} TS value of the dark matter component in the Virgo cluster for each $m_\chi$ and decaying channel. \emph{\textbf{Bottom.}} Reduction of the TS value of M87 for the corresponding $m_\chi$ and decaying channel.}
    \label{fig:ts_virgo}
\end{figure}

\begin{figure}[tbp]
\centering 
\includegraphics[width=.32\textwidth]{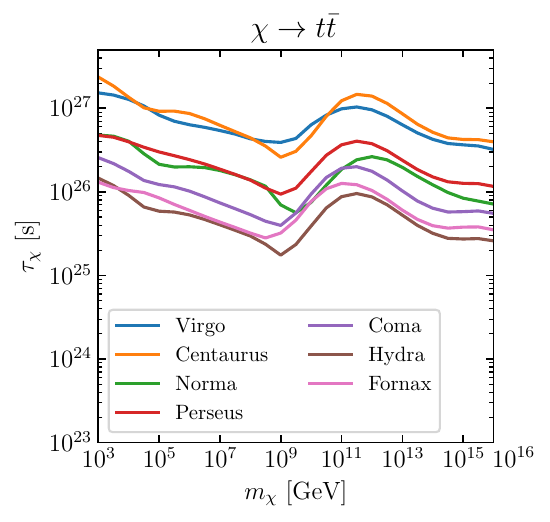}
\includegraphics[width=.32\textwidth]{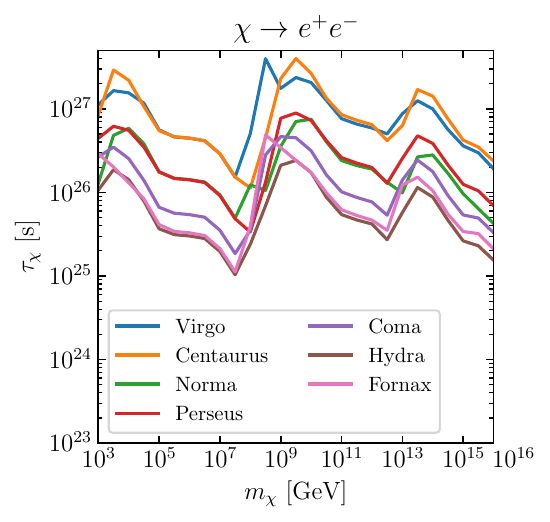}
\includegraphics[width=.32\textwidth]{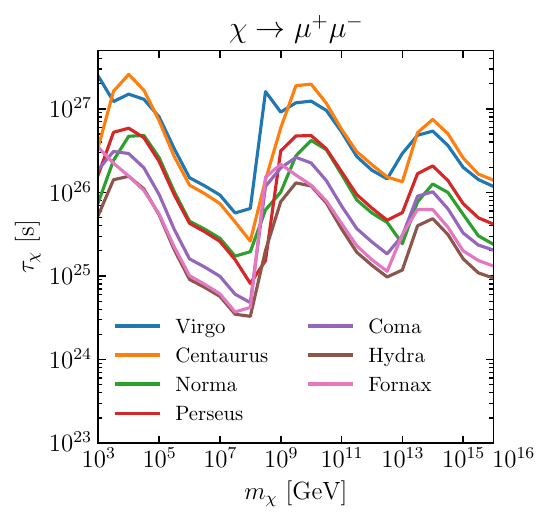}
\includegraphics[width=.32\textwidth]{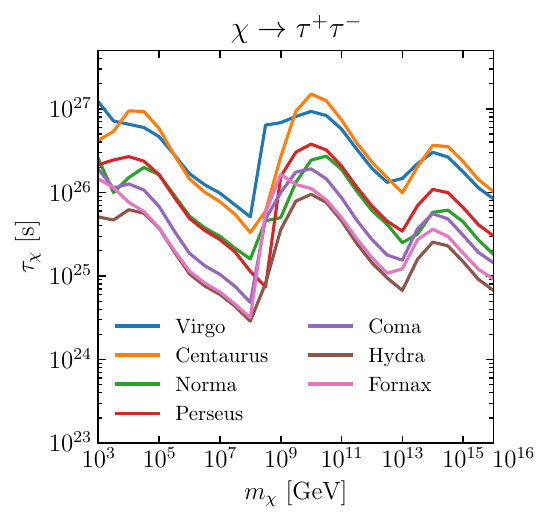}
\includegraphics[width=.32\textwidth]{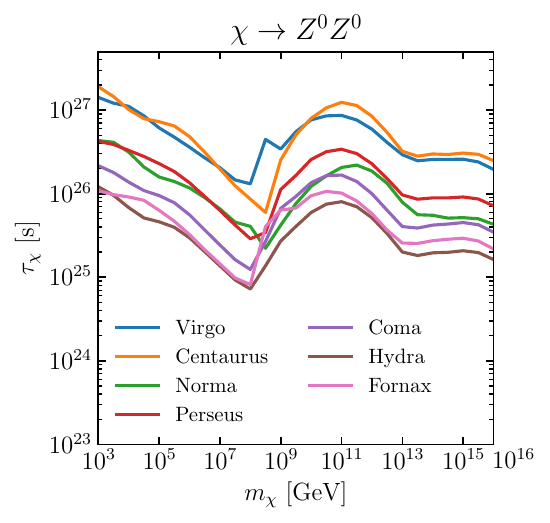}
\includegraphics[width=.32\textwidth]{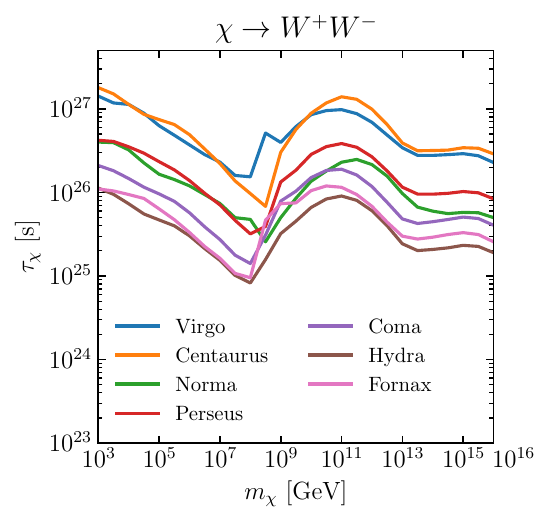}
\caption{\label{fig:limits} Limits on $\tau_\chi$ for the other decaying channels: $t\bar{t}$, $e^+e^-$, $\mu^+\mu^-$, $\tau^+\tau^-$, $Z^0Z^0$, and $W^+W^-$.}
\end{figure}

Figure~\ref{fig:limits_bbar} shows the limits for the $b\bar{b}$ channel from seven galaxy clusters. Generally, larger $J^\mathrm{dec.}$ in clusters provide more stringent limits, as expected. Across the entire mass range, the most stringent limits are set by two clusters: Virgo and Centaurus. The Virgo cluster has a $J^\mathrm{dec.}$ nearly one order of magnitude larger than that of Centaurus. However, the limits set by Centaurus are overall comparable to and often stronger than Virgo. This is largely caused by the fact that the Messier 87 (M87) galaxy at the center of Virgo cluster is detected as an active galactic nucleus by {\it Fermi}-LAT~\cite{2009ApJ...707...55A}. Since we refit the background sources when we derive the limits, a fraction of the gamma-ray flux from M87 is attributed to dark matter in the Virgo cluster due to their spatial overlapping, leading to weakened limits on $\tau_\chi$. On the top panel of figure~\ref{fig:ts_virgo}, we show TS$_\mathrm{Virgo}$, the TS value of the dark matter component in the Virgo cluster for each $m_\chi$ and decaying channel. In most cases, TS$_\mathrm{Virgo}$ are less than 1. On the bottom panel, we show $-\Delta\mathrm{TS}_\mathrm{M87}$, the reduction of the TS value of M87 when the corresponding dark matter model is included in the fit. It is obvious that $-\Delta\mathrm{TS}_\mathrm{M87}$ correlates with TS$_\mathrm{Virgo}$. This clearly shows how M87 affects the inferred dark matter limits from the Virgo cluster. It is known that background sources are impactful for dark matter constraints in general. By refitting background sources in all clusters, we obtain conservative and robust limits. Figure~\ref{fig:limits} shows the limits for the remaining decaying channels. Same as the $b\bar{b}$ channel, the Virgo and Centaurus clusters together set the most stringent constraints on $\tau_\chi$.


\begin{figure}[tbp]
\centering 
\includegraphics[width=.49\textwidth]{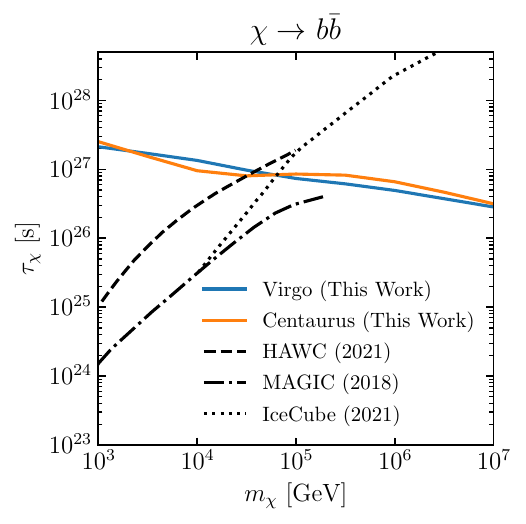}
\includegraphics[width=.49\textwidth]{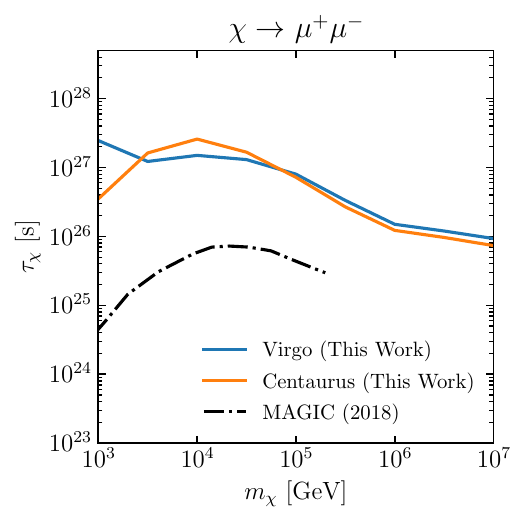}
\includegraphics[width=.49\textwidth]{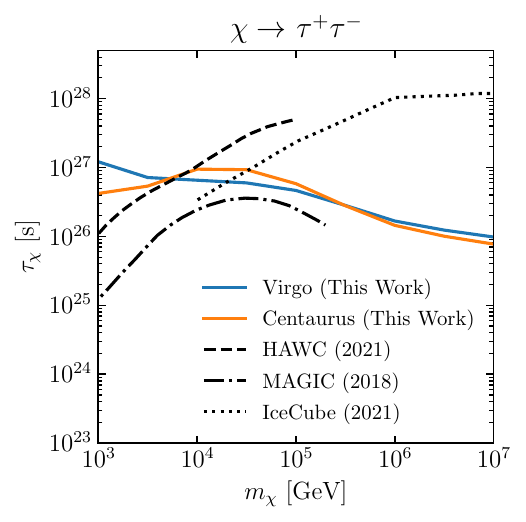}
\includegraphics[width=.49\textwidth]{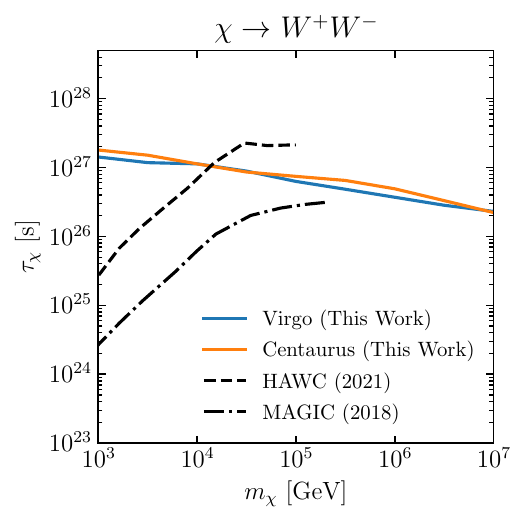}
\caption{\label{fig:limits_compare} Comparisons between our limits and recent gamma-ray (MAGIC~\cite{MAGIC:2018tuz} and HAWC~\cite{HAWC:2021udn}) and neutrino (IceCube~\cite{IceCube:2021sog}) observations on galaxy clusters.}
\end{figure}

In figure~\ref{fig:limits_compare}, we compare our results with limits set by galaxy clusters from recent gamma-ray and neutrino observations. For decaying dark matter, MAGIC has reported limits on $\tau_\chi$ in the mass range between 200 GeV and 200 TeV from the Perseus cluster using 400 hours of data between 2009 and 2017~\cite{MAGIC:2018tuz}. HAWC has also performed a search from the Virgo cluster for dark matter masses between 1 TeV and 100 TeV using 1523 days of observation~\cite{HAWC:2021udn}. Using 6-year neutrino data, IceCube has constrained decaying dark matter in the mass range between 10 TeV and 10 PeV from observations of three clusters (Virgo, Coma, and Perseus)~\cite{IceCube:2021sog}. In figure~\ref{fig:limits_compare}, we show the HAWC limits in dashed lines, the MAGIC limits in dashed-dotted lines, and the IceCube limits, which we have converted from 90\% CL to 95\% CL according to the $\chi^2$ distribution, in dotted lines. We compare these limits with our results from Virgo (blue solid lines) and Centaurus (orange solid lines) since they set the most stringent limits. We compare the limits for $10^3\ \mathrm{GeV} < m_\chi < 10^7\ \mathrm{GeV}$. Our limits are more stringent than MAGIC for the four available channels ($b\bar{b}$, $\mu^+\mu^-$, $\tau^+\tau^-$, and $W^+W^-$). At $m_\chi = 10^3$ GeV, our limits are stronger than MAGIC's by about 2 to 3 orders of magnitude since {\it Fermi}-LAT can fully cover the prompt emission from dark matter. HAWC limits are not available for the $\mu^+\mu^-$ channel. For other channels, HAWC limits are generally stronger than MAGIC by one order of magnitude. Our limits are usually stronger than HAWC for $m_\chi\lesssim 10^4$ GeV. We note that limits from MAGIC and HAWC do not consider the electromagnetic cascade from VHDM decay. IceCube limits are available for the $b\bar{b}$ and $\tau^+\tau^-$ channels. Our limits are competitive with IceCube for $m_\chi\lesssim 10^5$ GeV. However, IceCube can access much higher energies, and therefore has set the strongest limits for $m_\chi > 10^5$ GeV.

\subsection{Combined analysis}
So far, we have reported limits on $\tau_\chi$ from individual clusters. It is also possible to combine the analyses and define the total likelihood function of all seven clusters:
\begin{equation}
    \mathcal{L}(\theta) = \prod_k\mathcal{L}^k(\theta).
\end{equation}
We derive the combined limits at 95\% CL at
\begin{equation}
    {-2\Delta\log \mathcal{L}= -2\left(\log\mathcal{L}(\hat{\theta}) - \log\mathcal{L}(\theta) \right) = -2.71.}
\end{equation}
Naturally, $\tau_\chi$ is assumed to be the same in all clusters and the likelihood functions are combined. The background parameters $\lambda_l$ now include all the point sources and diffuse emissions in the ROIs of the clusters. Although we use the same models for the diffuse emissions, their parameters are allowed to have different values for different clusters.

\begin{figure}[tbp]
    \centering
    \begin{minipage}[t]{0.49\textwidth}
        \centering
        \includegraphics[width=\textwidth]{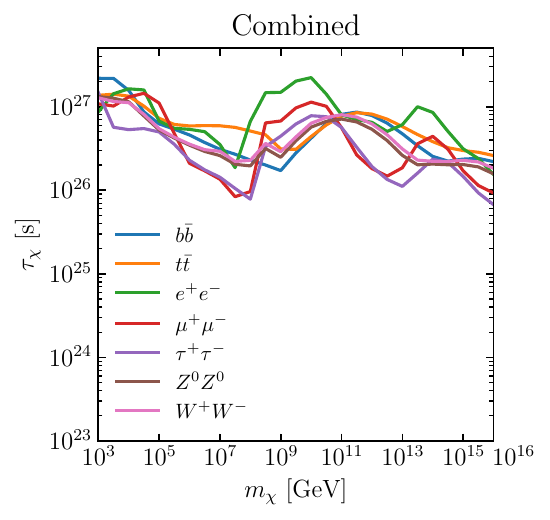}
        \caption{Limits on $\tau_\chi$ for five decaying channels from combined analyese of seven clusters.}
        \label{fig:limits_combined}
    \end{minipage}
    \vspace{1.00mm}
    \begin{minipage}[t]{0.49\textwidth}
        \centering
        \includegraphics[width=\textwidth]{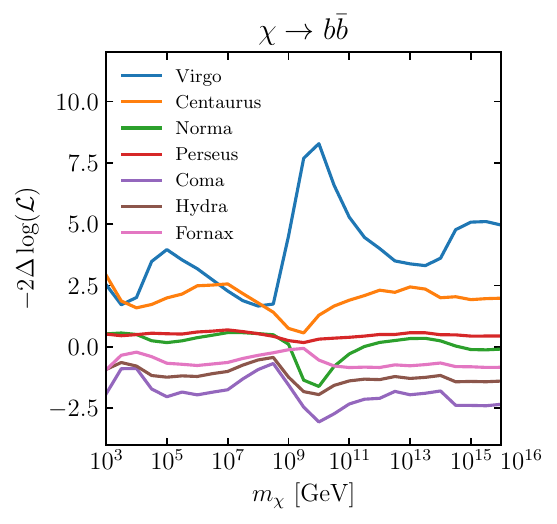}
        \caption{Contributions to $-2\Delta\log(\mathcal{L})$ at the 95\% CL limits for the $b\bar{b}$ channel from seven clusters in the combined analyses.}
        \label{fig:dloglike_combined_channel_5}
    \end{minipage}
\end{figure}

Figure~\ref{fig:limits_combined} shows the combined limits from all seven clusters. Overall, they are slightly weaker than the strongest limits set by Virgo and Centaurus individually. This is driven by clusters with small $J^\mathrm{dec.}$. Figure~\ref{fig:dloglike_combined_channel_5} shows the contributions to $-2\Delta\log(\mathcal{L})$ from the seven clusters when the 95\% CL limits on $\tau_\chi$ are set for the $b\bar{b}$ channel and for different $m_\chi$. Fornax, Hydra, Coma, and partially Norma contribute negatively to the $-2\Delta\log(\mathcal{L})$, meaning that including them in the combined analyses weakens the limits. This is due to their small expected signals and potentially more complicated background environments (e.g., Norma is located close to the Galactic plane). Previous study also finds similar results in combined likelihood analysis~\cite{Huang:2011xr}. Therefore, we find that promising individual clusters provide the strongest probe of VHDM.

\subsection{Systematic uncertainties}
\begin{figure}[tbp]
    \centering
    \begin{minipage}[t]{0.49\textwidth}
        \centering
        \includegraphics[width=\textwidth]{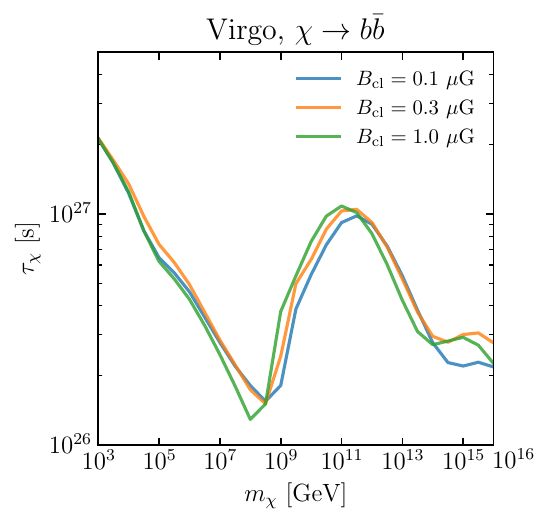}
        \caption{Limits on $\tau_\chi$ from Virgo for the $b\bar{b}$ channel assuming different $B_\mathrm{cl}$.}
        \label{fig:limit_bfield}
    \end{minipage}
    \vspace{1.00mm}
    \begin{minipage}[t]{0.49\textwidth}
        \centering
        \includegraphics[width=\textwidth]{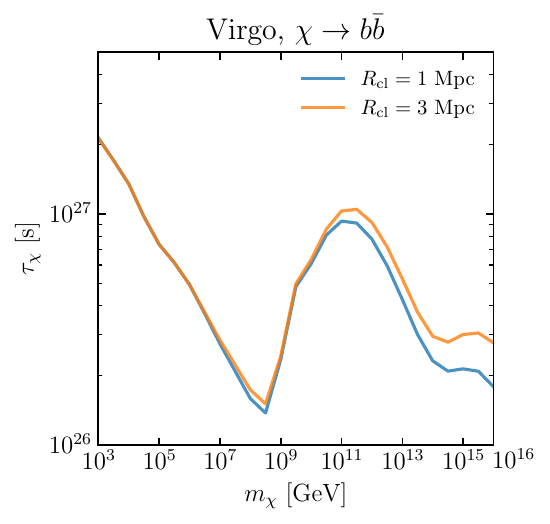}
        \caption{Limits on $\tau_\chi$ from Virgo for the $b\bar{b}$ channel assuming different $R_{\rm cl}$.}
        \label{fig:limit_size}
    \end{minipage}
\end{figure}

To derive limits on $\tau_\chi$, we take into account the electromagnetic cascades of prompt gamma rays and $e^\pm$ from dark matter decay. Therefore, we have to consider systematic uncertainties raised from the calculation of the cascade flux. Two factors are mostly influential: the magnetic field $B_\mathrm{cl}$ in the cluster in which $e^\pm$ lose energies and the escaping distance $R_{\rm cl}$ at which gamma rays leave the cluster and enter the intergalactic space. We have adopted $B_\mathrm{cl} = 0.3\ \mu$G and $R_{\rm cl} = 3$ Mpc. Typical intracluster magnetic fields in galaxy clusters average to $\sim 0.1-1\ \mu$G~\cite{2004IJMPD..13.1549G}. For the seven clusters in this work, their virial radii are $\sim$ 1 -- 3 Mpc (see table~\ref{tab:clusters}). In figure~\ref{fig:limit_bfield}, we show the limits from Virgo for the $b\bar{b}$ channel by using $B_\mathrm{cl} = 0.1, 0.3,\ \mathrm{and}\ 1.0\ \mu$G. The limits are largely the same at lower masses when the prompt emission is more important. The limits slightly change shapes for $m_\chi \gtrsim 10^5$ GeV but maintain roughly the same level.
{Realistically, the magnetic fields within the volume of the galaxy clusters vary, and the spectral features of the VHDM signals would alter correspondingly. Nonetheless, the constraints are not expected to dramatically change, as they are primarily determined by the predicted fluxes in the \textit{Fermi}-LAT energy range and electromagnetic cascades smear out the spectra. Therefore, the impacts of varying magnetic fields within the volume of the clusters are expected to lie around the range of the constraints shown in figure~\ref{fig:limit_bfield}.}
Figure~\ref{fig:limit_size} shows the limits from Virgo for the $b\bar{b}$ channel, using $R_{\rm cl} =$ 1 and 3 Mpc. For $m_\chi \gtrsim 10^{10}$ GeV, the limits from $R_{\rm cl} = 3$ Mpc are slightly stronger since stronger synchrotron emission is expected from larger $R_{\rm cl}$. However, the effect is not significant. We expect similar uncertainties for different clusters and decay channels. Overall, our limits are robust when $B_\mathrm{cl}$ and $R_{\rm cl}$ are in reasonable ranges.

\section{Summary}\label{sec:summary}
We searched for gamma-ray signals from very heavy dark matter decay from nearby galaxy clusters using 14 year {\it Fermi}-LAT data and set lower limits on the lifetime of decaying dark matter as a function of dark matter mass. Notably, we considered gamma rays from both the prompt emission and the electromagnetic cascades in the clusters. The cascade component is essential for constraints at large masses. Seven decaying channels are included: $b\Bar{b}$, $t\Bar{t}$, $e^+e^-$, $\mu^+\mu^-$, $\tau^+\tau^-$, $Z^0Z^0$, and $W^+W^-$. We use the profile likelihood method to obtain the limits for seven galaxy clusters: Virgo, Centaurus, Norma, Perseus, Coma, Hydra, and Fornax. We found that gamma-ray observations in the direction of Virgo and Centaurus clusters provide the most stringent limits on dark matter lifetime. We also reported the limits from combined analyses of seven clusters. Our limits remain strong and 
robust when we consider different magnetic field strengths and sizes of the clusters. Our results are competitive with limits from other gamma-ray observations on galaxy clusters and are complementary to previous gamma-ray and neutrino constraints on decaying very heavy dark matter.

\acknowledgments
The authors thank the anonymous referee for their valuable comments and suggestions. We thank Saikat Das, Nagisa Hiroshima, and B. Thedore Zhang for fruitful discussions. D.S. is supported by KAKENHI No. 20H05852. 
The work of K.M. was supported by the NSF Grants No.~AST-2108466 and No.~AST-2108467, and KAKENHI No.~20H01901 and No.~20H05852. A.K. is supported by the NASA grant 80NSSC23M0104.
This work uses the computational resources provided by the super computer Yukawa-21 at Yukawa Institute for Theoretical Physics.

% \paragraph{Note added.} This is also a good position for notes added
% after the paper has been written.

\appendix
\section{{Electromagnetic cascades inside galaxy clusters}}
\label{sec:cascades}
Following Ref.~\cite{Murase:2012rd} (see also Refs.~\cite{Murase:2008yt,Murase:2009zz,Kotera:2009ms} in the context of cosmic rays) , our numerical code solves the following Boltzmann equations in a time-dependent manner,
% \begin{equation}
\begin{align}
\frac{\partial N_\gamma(E_\gamma)}{\partial t}=&-N_\gamma \int d \varepsilon \frac{d n}{d \varepsilon} \int \frac{d \mu}{2} (1-\mu)c \sigma_{\gamma \gamma}(\varepsilon, \mu) -\frac{N_\gamma}{t_{\mathrm{esc}}}\\ \nonumber
&+\int d E^{\prime} N_e\left(E^{\prime}\right) \int d \varepsilon \frac{d n}{d \varepsilon} \int \frac{d \mu}{2} (1-\mu)c \frac{d \sigma_{\mathrm{IC}}}{d E_\gamma}\left(\varepsilon, \mu, E^{\prime}\right)\\ \nonumber
&+\frac{\partial N_\gamma^{\mathrm{syn}}}{\partial t}+Q_\gamma^{\mathrm{inj}}, \\
\frac{\partial N_e(E_e)}{\partial t}=&-N_e \int d \varepsilon \frac{d n}{d \varepsilon} \int \frac{d \mu}{2} (1-\mu)c \sigma_{\mathrm{IC}}(\varepsilon, \mu)\\\nonumber
&+\int d E^{\prime} N_\gamma\left(E^{\prime}\right) \int d \varepsilon \frac{d n}{d \varepsilon} \int \frac{d \mu}{2} (1-\mu)c \frac{d \sigma_{\gamma \gamma}}{d E_e}\left(\varepsilon, \mu, E^{\prime}\right)\\\nonumber
&+\int d E^{\prime} N_e\left(E^{\prime}\right) \int d \varepsilon \frac{d n}{d \varepsilon} \int \frac{d \mu}{2} (1-\mu)c \frac{d \sigma_{\mathrm{IC}}}{d E_e}\left(\varepsilon, \mu, E^{\prime}\right)\\\nonumber
&-\frac{\partial}{\partial E}\left[P_{\mathrm{syn}} N_e\right]+Q_e^{\mathrm{inj}}.
\end{align}
% \end{equation}
The background photon number density is $dn/d\varepsilon$ at the energy $\varepsilon$. For infrared and optical photons in galaxy clusters, we adopt the low-IR model of the EBL light from Ref.~\cite{Kneiske:2003tx} with a 10 times enhancement to account for contributions from the galaxies in the clusters~\cite{Silva:1998zz, 2012ApJ...748....9T, Murase:2012rd}. This is reasonable within uncertainties (see Refs.~\cite{2012ApJ...748....9T, Kotera:2009ms,Fang:2017zjf}).
The CMB is also included. We have implemented the accurate cross sections for pair production ($\sigma_{\gamma\gamma}$) and inverse-Compton scattering ($\sigma_\mathrm{IC}$) including the Klein-Nishina effect, as in calculations of intergalactic propagation of gamma rays and cosmic rays~\cite{Murase:2011cy, Murase:2011yw, Murase:2012df, Murase:2013rfa}.
Here, $P_\mathrm{syn}$ is the energy loss rate due to the synchrotron radiation. The differential synchrotron radiation spectrum $\partial N_\gamma^{\mathrm{syn}}/{\partial t}$ is 
\begin{align}
    \frac{\partial N_\gamma^{\mathrm{syn}}(E_\gamma)}{\partial t} &= \int d E^{\prime} N_e\left(E^{\prime}\right)\frac{\sqrt{3} e^3 B}{m_e c^2 2 \pi \hbar E_\gamma} G(x), \\
    G(x) &\approx \frac{1.81 e^{-x}}{\left(x^{-2 / 3}+(3.62 / \pi)^2\right)^{1 / 2}},
\end{align}
where $x = E_\gamma/E_c$ and $E_c$ is the critical energy. The function form $G(x)$ is an accurate fit to the synchrotron radiation spectrum~\cite{Zhang:2020qbt,2010PhRvD..82d3002A}.
% Accurate formulations for the pair production, inverse-Compton emission (including the Klein-Nishina effect), and the synchrotron emission are considered. 
The diffusion of electrons and positrons is negligible for high-energy radiation~\cite{Murase:2011yw}. 
For example, at TeV energies, the $e^\pm$ cooling time is $t_\mathrm{cool} \approx t_\mathrm{IC}\sim 1\ \mathrm{Myr}\ \left(\dfrac{{\rm TeV}}{E_e}\right) \left(\dfrac{u_{\rm CMB}}{u_{\rm rad}}\right)$~\cite{Murase:2011yw}. 
This is much shorter than the light crossing time, $t_\mathrm{esc} = R_\mathrm{cl}/c\sim 10\ \mathrm{Myr} \left(\dfrac{R_\mathrm{cl}}{3\ \mathrm{Mpc}} \right)$. 
The diffusion time scale $t_\mathrm{diff}$ is always longer than the light crossing time $t_\mathrm{esc}$, implying $t_\mathrm{cool}\ll t_\mathrm{diff}$, which is also the case in the calculations of cascade gamma rays induced by cosmic rays confined in galaxy clusters~\cite{Murase:2009zz,Murase:2012rd}. 

The Boltzmann equations are numerically binned in finite energy and time steps and are solved iteratively to obtain essentially steady-state spectra for high-energy gamma rays~\cite{Murase:2012rd}.

\bibliographystyle{jhep}
\bibliography{main}


% \begin{thebibliography}{99}

% \bibitem{a}
% Author, \emph{Title}, \emph{J. Abbrev.} {\bf vol} (year) pg.

% \bibitem{b}
% Author, \emph{Title},
% arxiv:1234.5678.

% \bibitem{c}
% Author, \emph{Title},
% Publisher (year).

% Please avoid comments such as "For a review'', "For some examples",
% "and references therein" or move them in the text. In general,
% please leave only references in the bibliography and move all
% accessory text in footnotes.

% Also, please have only one work for each \bibitem.

% \end{thebibliography}
\end{document}